\documentstyle[referee]{laa}

\begin{document}
\thesaurus{07  
	   (07.09.1; 
	    07.13.1;  
	   )}

\title{Orbital Motion in Outer Solar System}
\author{J.~Kla\v{c}ka $^{1)}$ and M.~Gajdo\v{s}\'{\i}k $^{2)}$}
\institute{Institute of Astronomy,
   Faculty for Mathematics and Physics, Comenius University \\
   Mlynsk\'{a} dolina, 842~15 Bratislava, Slovak Republic $^{1)}$ \\
   Ostredsk\'{a} 20,
   821~02 Bratislava, Slovak Republic $^{2)}$}
\date{}
\maketitle

\begin{abstract}
Motion of a point mass in gravitational fields of the Sun and of
the galactic disk is studied. Fundamental features of the motion are
found by investigating the time-averaged differential equations for
orbital evolution.
Several types of possible orbits
are mathematically exactly derived
in a strictly analytical way. The relation
$a^{3} ~ P^{2} = f ( e_{0}, i_{0}, \omega_{0} )$ between semimajor axis $a$
and period $P$ of the change of osculating orbital elements is found
(the index $0$ denotes initial values of the quantities).

Due to conservation of energy in potential fields $a$ is a constant.
Moreover, the component of angular momentum perpendicular to the
galactic plane is conserved. Due to these facts the system of equations
reduces to two equations for either ($e$, $\omega$), or ($i$, $\omega$)
(the length of the ascending node does not enter the equations for
$a$, $e$, $i$, $\omega$ and is not solved here).


\end{abstract}

\section{Introduction}
Motion of a point mass in gravitational fields of the Sun and of
the galactic disk is important from the point of view of orbital
evolution of cometary orbits. One of such numerical studies
may be found, e. g., in Pretka and Dybczynski (1994).
Since we want to find fundamental features of
possible motions, we use time-averaged equations for orbital elements
($e$, $i$, $\omega$) --
conclusions may be applied to real situations for semimajor axes
up to the order of 10$^{4}$ AU.


\section{Dynamical Model}
The dynamical model discussed in this paper is given by the following equations
of motion:
\begin{eqnarray}\label{1}
\ddot{x} &=& - ~ \frac{\mu}{r^{3}} ~ x \nonumber \\
\ddot{y} &=& - ~ \frac{\mu}{r^{3}} ~ y \nonumber \\
\ddot{z} &=& - ~ \frac{\mu}{r^{3}} ~ z	~-~ k~ z ~,
\end{eqnarray}
where $\mu = G ~M_{\odot}$, $k = 4 ~\pi ~G ~\varrho$ and $\varrho$ is the mean
density of the mass of the galactic disk in the Solar neighbourhood.

\section{Time-averaged Equations for Orbital Elements}
The case $k = 0$ in Eqs. (1) corresponds to a Keplerian motion. If $k$ is smaller
than  $\mu ~/~ r^{3}$, we can consider the term $k$ $z$ as a disturbing
acceleration to two-body problem. The disturbing acceleration is given
by
\begin{equation}\label{2}
\vec{a} = ( 0, 0, -~ k ~z ) = ( 0, 0, - ~ k ~ r ~ \sin i ~ \sin \Theta )
\end{equation}
in rectangular coordinates, where $\Theta = \omega ~+~ f$, $f$ is the
true anomaly.
Radial, normal and transversal components of the disturbing acceleration
are:
\begin{eqnarray}\label{3}
a_{R} &=& \vec{a} \cdot \hat{e}_{R} =
	 - ~ k ~ r~ ( \sin i )^{2} ~ ( \sin \Theta )^{2}  \nonumber \\
a_{N} &=& \vec{a} \cdot \hat{e}_{N} =
	 - ~ k ~ r~ ( \sin i ) ~ ( \cos i )~ ( \sin \Theta )  \nonumber \\
a_{T} &=& \vec{a} \cdot \hat{e}_{T} =
	 - ~ k ~ r~ ( \sin i )^{2} ~ ( \sin \Theta ) ~ ( \cos \Theta ) ~.
\end{eqnarray}

Eqs. (3) enable us to obtain time-averaged evolutionary equations for
orbital elements. The significant equations are ($\Omega$ and $T$ do not enter
this system, and, in this paper, we are not interested in their evolution):
\begin{eqnarray}\label{4}
< \frac{d a}{d t} > &=& 0    \nonumber \\
< \frac{d e}{d t} > &=& \frac{5}{4} ~ k ~
			\sqrt{\frac{a^{3}}{\mu}} ~
			\left ( \sin i \right ) ^{2} ~
			\left [ \sin \left ( 2 ~ \omega \right ) \right ] ~
			e~ \sqrt{1 ~-~ e^{2}}	 \nonumber \\
< \frac{d i}{d t} > &=& -~ \frac{5}{8} ~ k ~
			\sqrt{\frac{a^{3}}{\mu}} ~
			\left [ \sin \left ( 2 ~i \right ) \right ] ~
			\left [ \sin \left ( 2 ~ \omega \right ) \right ] ~
			\frac{e^{2}}{\sqrt{1 ~-~ e^{2}}}    \nonumber \\
< \frac{d \omega}{d t} > &=& -~ \frac{5}{2} ~ k ~
			\sqrt{\frac{a^{3}}{\mu}} ~ \left \{
	 \frac{\left ( \sin i  \right ) ^{2} ~ - ~ e^{2}}{\sqrt{1 ~-~ e^{2}}} ~
	 \left ( \sin \omega \right ) ^{2} ~-~ \frac{1}{5} ~
			\sqrt{1 ~-~ e^{2}} \right \} ~.
\end{eqnarray}
The first of Eqs. (4) accords with conservation of energy in the system.

Eqs. (4) show that it is useful to define a dimensionless variable $\tau$:
\begin{equation}\label{5}
\tau \equiv \frac{5}{2} ~ k ~ \sqrt{\frac{a^{3}}{\mu}} ~ t ~.
\end{equation}
Moreover, equations for eccentricity and inclination define the relation
between these two quantities:
\begin{equation}\label{6}
\sqrt{1 ~-~ e^{2}} ~ \cos i =  \sqrt{1 ~-~ e_{0}^{2}} ~ \cos i_{0} ~,
\end{equation}
where the index $0$ denotes initial quantities. Eq. (6) corresponds
to conservation of the $z-$ component -- the component perpendicular to the
galactic plane -- of angular momentum together with the
condition $a = a_{0}$: $d \vec{H} / d t$ $=$ $\vec{r} \times a$ $=$
$\vec{r} \times$ ( $- \mu \vec{r} / r^{3} ~-~ k~z ~ \vec{e}_{N}$) $=$
$- \vec{r} \times \vec{e}_{N} ~k~z $ $\Rightarrow$ $H_{z} \equiv H_{N} =$
constant.

Eqs. (5) and (6) reduce Eqs. (4) to two equations:
\begin{eqnarray}\label{7}
< \frac{d i}{d \tau} > &=&  \frac{1}{2} ~
			\left [ \sin \left ( 2 ~ \omega \right ) \right ] ~
			\left ( \sin i \right ) ~
\frac{\left ( 1~- e_{0}^{2} \right ) ~\left ( \cos i_{0} \right )^{2} ~-~
\left ( \cos i \right )^{2}}{\sqrt{1 ~-~ e_{0}^{2}} ~ \cos i_{0}}    \nonumber \\
< \frac{d \omega}{d \tau} > &=&
\frac{\left [ 1 / 5 ~-~ \left ( \sin \omega  \right ) ^{2} \right ] ~
\left ( 1~- e_{0}^{2} \right ) ~\left ( \cos i_{0} \right )^{2} ~+~
\left ( \sin \omega  \right ) ^{2} ~\left ( \cos i \right )^{4}}{
	  \sqrt{1 ~-~ e_{0}^{2}} ~
	 \left ( \cos i_{0} \right ) ~
	 \left ( \cos i \right ) } ~.
\end{eqnarray}

\section{Mathematical Treatment}
Eqs. (7) form a complete set of differential equations -- initial conditions
$\omega_{0}$, $i_{0}$ must be added, of course. We may interpret solutions
of Eqs. (7) in terms of several types
of orbits.

At first
\begin{equation}\label{8}
\left ( \cos i_{0} \right ) ^{2} = \frac{4}{5} ~
				   \left ( 1~-~ e_{0}^{2} \right )  ~\wedge ~
\left ( \omega_{0} = \frac{\pi}{2} ~\vee ~ \omega_{0} = \frac{3~ \pi}{2}
\right ) \Longrightarrow ~
	 < \frac{d i}{d \tau} > ~ = ~ < \frac{d \omega}{d \tau} >  ~ = 0 ~.
\end{equation}

Eq. (8) define one type of orbits -- stationary orbits. The remaining types
correspond to periodic changes of the time-averaged elements.

When considering periodic changes, one would await that $\dot{\omega}$
may be positive as well as negative. One obtains, on the basis of the
second of Eqs. (7)
\begin{equation}\label{9}
\dot{\omega} < 0  \Longleftrightarrow
\frac{1}{5} ~ \left ( 1~-~ e^{2} \right ) ~   <
\left ( \sin \omega ~ \sin i \right ) ^{2} ~ \left \{
      1 ~-~ \left ( \frac{e}{\sin i} \right ) ^{2} \right \} ~.
\end{equation}
Eq. (9) states that if the case $\dot{\omega} < 0$ may occur, the quantity
$\omega$ can change periodically only in a small interval given by the values
of $e$ and $i$ -- oscillations in $\omega$;
the same holds for the quantity $\sin b = \sin \omega ~ \sin i$. This is
the second type of orbits. (The consequence of Eq. (9) is
$e < \sin i$, and,
$\left ( \sin b \right ) ^{2} > 1/5$.)

The final type of orbits corresponds to the case when the condition
$\dot{\omega} > 0$ is permanently fulfilled: relation (9) can never hold during the
orbital motion.
In this case
$\omega$ increases monotonically from 0 to 2~$\pi$, $b$ oscillates
about zero.

The last two cases should correspond to the types of orbits found in
Pretka and Dybczynski (1994) in a numerical way -- by numerical integration
of Eqs. (1). The two types of possible orbits are separated by the
conditions:  i) always $\dot{\omega} >$ 0 holds
($\left ( \sin b \right ) ^{2} <$ 1/5 $\Leftrightarrow$ $\dot{\omega} >$ 0,
always);
ii) oscillations in $\omega$
(oscillations in $\omega$ $\Leftrightarrow$
$\left ( \sin b \right ) ^{2} > 1/5$).
We see that the results do not
depend on semimajor axis and eccentricity (the statement in
Pretka and Dybczynski (1994)), but on eccentricity $e$ and $e / \sin i$.

\section{One consequence}
The consequence of Eq. (5) and periodical solutions of Eqs. (7) is that
the relation
$a^{3} ~ P^{2} = f ( e_{0}, i_{0}, \omega_{0} )$ between semimajor axis $a$
and period $P$ of the change of the time-averaged orbital elements exists.
This relation is interesting in comparison with the Kepler's third law:
$a^{3} ~/~ T^{2} = \mu ~/~ ( 4 ~ \pi ^{2} )$ ... $T$ is the period
of revolution in the ellipse.

\section{Conclusion}
We have found, on a firm mathematical basis, all types of orbits
which exist in the outer part of the Solar System for a given model.
Three types of orbits exist: i) always $< \dot{\omega} > = 0$
(the same holds for other orbital elements),
ii) $< \dot{\omega} >$ is always positive, and,
iii) $< \omega >$ oscillates during the orbital motion.
We have
completed and put into a correct form the ``experimental'' numerical
results of Pretka and Dybczynski (1994), all in an analytical
way. An interesting result
$a^{3} ~ P^{2} = f ( e_{0}, i_{0}, \omega_{0} )$ between semimajor axis $a$
and period $P$ of the change of time-averaged orbital elements was obtained.

\acknowledgements
The work of one of the authors (J. K.) was partially supported by the
Scientific Grant Agency VEGA
(grants Nos. 1/4304/97 and 1/4303/97).

\end{document}